# Beltrami-like fields created by baroclinic effect in two-fluid plasmas

H. Saleem
*PINSTECH (NDP), P. O. Nilore, Islamabad, Pakistan, and COMSATS Institute of Information Technology (CIIT), Department of Physics H-8, Islamabad, Pakistan*

Z. Yoshida
*University of Tokyo, Graduate School of Frontier Sciences, Tokyo 113-0033, Japan*

A theory of two-dimensional plasma evolution with Beltrami-like flow and field due to baroclinic effect has been presented. Particular solution of the nonlinear two-fluid equations is obtained. This simple model can explain the generation of magnetic field without assuming the presence of a seed in the system. Coupled field and flow naturally grow together. The theory has been applied to estimate B-field in laser-induced plasmas and the result is in good agreement with experimental values. © *2004 American Institute of Physics.* [DOI: 10.1063/1.1793173]

The baroclinic effect, represented by the term $\nabla n \times \nabla T$ ($n$ represents density, and $T$ the temperature) in the magnetoplasma equations,[1–4] plays a key role in generating magnetic fields and coupled flows in inhomogeneous plasmas. Quantitative estimates of the magnetic field generation proves the effectiveness of this mechanism in laser-irradiated targets. In these pioneering works, however, the self-consistency of the flow and magnetic fields was not seriously addressed, because finding exact solutions to the determining nonlinear system of equations is generally rather difficult.

In this Brief Communication, we present rather simple, but nontrivial solutions to the nonlinear equations of two-fluid magnetohydrodynamics. We generalize the Beltrami conditions[5] with incorporating the baroclinic term, and derive a set of linear equations that give exact and self-consistent solutions to the full system of the nonlinear equations. The model predicts that density inhomogeneity set in perpendicular temperature gradient yields a growing magnetic field and simultaneous shear flow.

We assume a time scale that is much longer than the electron plasma oscillation time ($\omega_{pe}^{-1}$), and we ignore the electron inertia effect. The governing equations are[5]

$$\partial_t n_j + \nabla \cdot (\mathbf{v}_j n_j) = 0 \quad (j=e,i), \tag{1}$$

$$\partial_t \mathbf{A} - (\mathbf{v}_e \times \mathbf{B}) = -\nabla \Phi + \frac{\nabla p_e}{n_e}, \tag{2}$$

$$\partial_t (\mathbf{A} + \mathbf{v}_i) - \mathbf{v}_i \times (\mathbf{B} + \nabla \times \mathbf{v}_i) = -\nabla \left(\Phi + \frac{1}{2}v_i^2\right) - \frac{\nabla p_i}{n_i}, \tag{3}$$

where $n_i(n_e)$ is the ion (electron) density, $\mathbf{v}_i(\mathbf{v}_e)$ is the ion (electron) flow velocity (normalized to the Alfvén speed $v_A = B_0/\sqrt{4\pi M n_i}$, $\mathbf{B}$ is the magnetic field normalized to a representative value $B_0$), $p_i(p_e)$ is the ion (electron) pressure (normalized to the magnetic pressure $B_0^2/4\pi$), and $\mathbf{A}$ and $\Phi$ are the four potentials of the electromagnetic fields. The space and time are normalized to the ion skin depth $\lambda_i = c/\omega_{pi} = \sqrt{cM/(4\pi n_{i0}e^2)}$ and the ion cyclotron time $\omega_{ci}^{-1}$

$= Mc/(eB_0)$, respectively. Assuming quasineutrality ($n_i \approx n_e \approx n$) and singly charged ions, we have

$$\mathbf{j} = n(\mathbf{v}_i - \mathbf{v}_e). \tag{4}$$

We, thus, may write $\mathbf{v}_e = \mathbf{v}_i - (\mathbf{j}/n)$. For simplicity, we assume that the plasma is incompressible, i.e., $\nabla \cdot \mathbf{v}_j = 0$ ($j=e,i$).

Denoting $\psi = \ln n$ in Eq. (1) and taking the curl of Eqs. (2) and (3), the system of equations reads

$$\partial_t \psi + \nabla \cdot (\mathbf{v}_j \psi) = 0, \tag{5}$$

$$\partial_t \mathbf{B} + \nabla \times \{\mathbf{B} \times [\mathbf{v}_i - (\mathbf{j}/n)]\} = -\nabla \psi \times \nabla T_e, \tag{6}$$

$$\partial_t (\mathbf{B} + \nabla \times \mathbf{v}_i) - \nabla \times [\mathbf{v}_i \times (\mathbf{B} + \nabla \times \mathbf{v}_i)] = \nabla \psi \times \nabla T_i. \tag{7}$$

The reason for taking curl of Eqs. (2) and (3) becomes clear from Eqs. (6) and (7) where the sources for the plasma flow and magnetic field become the baroclinic terms on the right hand sides.

We consider a two-dimensional geometry where the densities and temperatures vary in the *x-y* plane. We can express the vector fields in Clebsch forms:

$$\mathbf{v}_i = \nabla \phi \times \mathbf{z} + u_i \mathbf{z}, \tag{8}$$

$$\mathbf{B} = \nabla \chi \times \mathbf{z} + \omega \mathbf{z}, \tag{9}$$

where the four scalar fields $\phi$, $u_i$, $\chi$, and $\omega$ are assumed to be independent of $z$. In the later calculations, we will assume that $u_i$ and $\chi$, as well as $\psi$ and $T_j$ ($j=i,e$), are constant with respect to $t$, while $\phi$ and $\omega$ may be function of $t$.

The continuity equations (5), with $\partial_t \psi = 0$, are satisfied, if

$$\{\phi, \psi\} = 0, \tag{10}$$

where $\{\varphi, \psi\} = \partial_y \varphi \partial_x \psi - \partial_x \varphi \partial_y \psi$.

Our aim is to find certain relations among physical quantities using Eqs. (8) and (9) in Eqs. (6) and (7) which can generate the coupled flow and field corresponding to given profiles of densities and temperatures. In our opinion, the baroclinic vectors can couple with the helicities of both magnetic field lines and streamlines in such a way that the non-

(10)

linear terms vanish in Eqs. (6) and (7). Then we can find exact solutions of the complicated nonlinear partial differential equations for given structures of densities and temperatures. For such an analysis we decompose the vector equation (7) into its three components.

The $x$ and $y$ components of the ion equation (7) give

$$\partial_t(\chi + u_i) + \{\phi,(\chi + u_i)\} = 0, \tag{11}$$

while the $z$ component of Eq. (7) yields

$$\partial_t(\omega - \Delta\phi) + [\{\phi,(\omega - \Delta\phi)\} + \{u_i,(\chi + u_i)\}] = -\{\psi,T_i\}. \tag{12}$$

The so-called "Beltrami" conditions simplify nonlinear vortex dynamics systems significantly, while the essential flow-vorticity coupling in the system is well highlighted in an analytically tractable form. We assume

$$\mathbf{B} = g(\nabla \times \mathbf{v}_i), \tag{13}$$

where $g$ is a constant number. Using the representations (8) and (9) in Eq. (13), we find

$$\omega = -g\Delta\phi, \tag{14}$$

$$\chi = gu_i. \tag{15}$$

We assume

$$\partial_t\chi = \partial_t u_i = 0. \tag{16}$$

Then, Eq. (11) holds, if

$$\{\phi,u_i\} = 0, \tag{17}$$

because Eq. (15) warrants $\{\phi,u_i\}=0$. Under the fourth Beltrami-like condition

$$\{\phi,\omega\} = 0, \tag{18}$$

the generalized vortex equation (12) reduces to

$$\partial_t(\omega - \Delta\phi) = \{T_i,\psi\}. \tag{19}$$

Integrating Eq. (19) with respect to $t$, we obtain growing $\phi$ and $\omega$. Let us assume, for simplicity, that $\psi$ and $T_i$ are independent of $t$. Using Eq. (14), we can integrate Eq. (19) to obtain

$$(g+1)^{-1}\omega = t\{T_i,\psi\}. \tag{20}$$

Now the determining equations for the ion-related fields are summarized as follows: Equation (14), together with Eq. (18), reads as a Poisson equation,

$$-\Delta\phi = g^{-1}\omega(\phi), \tag{21}$$

which must be consistent to Eqs. (10) and (20). The remaining scaler fields $u_i$ and $\chi$ are governed by the Beltrami conditions (15) and (17).

Let us preceed to formulate the electron Beltrami relations. Using the continuity equation (10), we observe

$$\nabla \times (\mathbf{v}_e \times \mathbf{B}) = \left[\{\chi,u_i\} + \{\omega,\phi\} + \left\{\chi,\frac{\Delta\chi}{n}\right\}\right]\mathbf{z}. \tag{22}$$

Here we assume

$$\{\chi,(\Delta\chi)/n\} = 0. \tag{23}$$

With the previous Beltrami conditions (10), (15), and (18), we find that the right hand side of Eq. (22) vanishes. The electron momentum equation (2) now reads

$$(\partial_t\omega)\mathbf{z} = -\nabla\psi_e \times \nabla T_e = -\{T_e,\psi\}\mathbf{z}, \tag{24}$$

which yields

$$\omega = -t\{T_e,\psi\}. \tag{25}$$

Let us derive an exact solution to the determining equations. Taking $g^{-1}\omega(\phi)=\lambda\phi$ ($\lambda$ is a real constant), Eq. (21) simplifies as a linear Poission equation:

$$\nabla\phi = -\lambda\phi. \tag{26}$$

The inhomogeneous term, representing the baroclinic effect, must be carefully chosen to obtain a set of self-consistent fields. Comparing Eqs. (10), (20), (25), and (26), we find that $T_j$ and $\psi$ must satisfy a relation

$$\{\{T_j,\psi\},\psi\} = 0 \quad (j=e,i).$$

Let us assume the temperatures to be given as

$$T_j = b_j y \quad (j=i,e), \tag{27}$$

and the density (common for ions and electrons) as

$$\psi = C\exp(\mu_1 x)\sin(\mu_2 y) \quad (j=i,e), \tag{28}$$

where $b_i, C, \mu_1$ and $\mu_2$ are arbitrary constants. The corresponding baroclinic term becomes

$$\{T_j,\psi\} = b_j\mu_1 C\exp(\mu_1 x)\sin(\mu_2 y) \quad (j=i,e).$$

Equations (20) and (25) yield, respectively,

$$\omega = (1+g)b_i\mu_1\psi \times t \tag{29}$$

$$= -b_e\mu_1\psi \times t. \tag{30}$$

The consistency demands

$$(1+g)b_i = -b_e. \tag{31}$$

Equation (26) is satisfied with $\lambda = \mu_2^2 - \mu_1^2$, and

$$\phi = \alpha b_i\mu_1\psi \times t, \tag{32}$$

where $\alpha$ is a constant. The condition (10) is also satisfied.

The simplest solution to the condition $\{\phi,u_i\}=0$ [see Eq. (17)] is given by, with a constant $U_0$,

$$u_i = U_0\alpha b_i\mu_1\psi. \tag{33}$$

Then, Eq. (15) gives

$$\chi = gU_0\alpha b_i\mu_1\psi. \tag{34}$$

The above example of self-consistent fields may capture the essential characteristics of structures that stem from inhomogeneous plasmas. Let us compare the solution with some examples of magnetic field and flow generations in laser-irradiated targets. In the experiments[6,7] and computer simulation studies[8] magnetic fields of the order of mega Gauss have been found to be produced in laser-induced plasmas.

Let $x$ be the vertical direction with respect to the surface of the pellet. We assume that the density has an exponential decay in $x$, as well as inhomogeneity in $y$, modeled by Eq. (28). We also assume that the temperature has gradients in the negative $y$ direction. To write the solution in physical units, we set $b_e = T_e/L_T$, $\mu_1 = 1/L_n$, and $C = n$, where $T_e$ is in eV, and $L_T$ and $L_n$ are electron temperature and density inhomogeneity scale lengths, respectively. Equation (32) in physical units can be written as

$$\omega = \frac{cT_e}{eL_T L_n} \sin(\mu_2 y) \exp(\mu_1 x) \times t. \quad (35)$$

The magnetic field and the flow have the following profiles:

$$\mathbf{v}_i = \begin{pmatrix} tP\mu_2 \exp(\mu_1 x)\cos(\mu_2 y) \\ -tP\mu_1 \exp(\mu_1 x)\sin(\mu_2 y) \\ U_0 P \exp(\mu_1 x)\sin(\mu_2 y) \end{pmatrix}, \quad (36)$$

$$\mathbf{B} = \begin{pmatrix} gU_0 P \exp(\mu_1 x)\cos(\mu_2 y) \\ -gU_0\mu_1 P \exp(\mu_1 x)\sin(\mu_2 y) \\ tQ \exp(\mu_1 x)\sin(\mu_2 y) \end{pmatrix}, \quad (37)$$

where $P = \alpha b_i \mu_i C$ and $Q = cT_e/(eL_n L_T)$.

Let us examine the magnitudes of the magnetic field and the flow. We use experimental parameters $n \approx 10^{22}$ cm$^{-3}$, $T_e \approx 1$ keV, $L_T \approx L_n = L \approx 0.005$ cm, and $t = L_n/c_s$ (the ion sound speed $c_s = 3 \times 10^7$ cm/sec). For $\omega_{pi}^{-1} \approx 7.6 \times 10^{-15}$ sec, we estimate the length scale $\lambda_i = 2.3 \times 10^{-4}$ cm, and hence $\lambda_i/L \approx 0.4$. Then, we obtain $tcT_e/(eL_T L_n) \approx 10^6$ G which is good agreement with observed values.[6]

In summary, we have derived a simple but interesting solution of nonlinear two-fluid plasma equations that captures some essential characteristics of the coupled flow magnetic fields generated by the baroclinic effect. The self-consistent fields are separated into growing and ambient (static) parts, and each part has different geometric characters—the toroidal (longitudinal) magnetic field $\omega$ and the poloidal flow $\phi$ grow simultaneously, while the poloidal magnetic field and toroidal flow are static. This is a natural consequence of the "canonical vorticity" (curl of the canonical momentum) defined in Eqs. (6) and (7). The baroclinic term is contravariant, which may couple only with the cononical vorticity consisting of $\omega$ and $\Delta\phi$. As a consequence of the "vorticity generation" due to the baroclinic effect, the helicities (twists) of both magnetic field lines and streamlines change gradually through the amplification, but they go differently.

We are predicting long-lived nonlinear structures that may stem in inhomogeneous plasmas. The instabilities of some ambient inhomogeneous plasma (the stationary fields in the model) may create magnetic fields and flows with specific forms due to the baroclinic term $\boldsymbol{\nabla}\psi \times \boldsymbol{\nabla}T_j$. The solution grows (not exponentially, but in proportion to $t$) from "zero," satisfying the nonlinear evolution equations exactly. Since the evolution equations are fully nonlinear consisting only second-order terms, and because the initial conditions may be zero, the solution is "stable" against small perturbation of the initial conditions. However, the behavior of the solution may change drastically, if we modify the ambient fields. It seems rather difficult to find other configurations where such a simple long-lived nonlinear structure may develop. When the created magnetic fileds and flows drive the system into turbulence and mess up the baroclinic term, no significant structures may be created.

This theoretical model has been applied to the laser-produced plasmas. The plasma ablates from the target surface when the laser pulse reaches it. The density increases in the direction perpendicular to the surface. The temperatures decrease away from the central laser hot spot on the plane of the target. Then, a magnetic field is generated in the $z$ direction at every point due to $\boldsymbol{\nabla}\psi \times \boldsymbol{\nabla}T_j$ which can circulate the laser beam axis. In this simple model the collisions have been ignored. Recently numerical simulation results based on collisional approach for the generation of magnetic field in laser-induced plasmas have also been presented.[9]

One of us (H.S.) thanks the University of Tokyo where this research started during his visit. The authors acknowledge the Abdus Salam ICTP where this research was completed during the Autumn College on Plasma Physics. They also acknowledge the discussions with Professor S. M. Mahajan.

[1] K. A. Brueckner and S. Jorna, Rev. Mod. Phys. **46**, 325 (1974).
[2] B. A. Altercop, E. V. Mishin, and A. A. Rukhadze, JETP Lett. **19**, 170 (1974 ).
[3] L. A. Bol'shov, A. M. Dykhne, N. G. Kowalski, and A. I. Yudin, in *Handbook of Plasma Physics*, edited by M. N. Rosenbluth and R. Z. Sagdeev, Physics of Laser Plasma, Vol. 3 (Elsevier Science, New York, 1991), pp. 519–548.
[4] A. A. Kingssep, K. V. Chukbar, and V. V. Yan'kov, in *Reviews of Plasma Physics*, edited by B. B. Kadomtsev (Consultants Bureau, New York, 1990), Vol. 16, p. 243.
[5] S. M. Mahajan and Z. Yoshida, Phys. Rev. Lett. **81**, 4863 (1998).
[6] J. A. Stamper, K. Papadopoulos, R. N. Sudan, S. O. Dean, E. A. Mclean, and J. W. Dawson, Phys. Rev. Lett. **26**, 1012 (1971).
[7] A. Raven, O. Willi, and P. T. Rumsby, Phys. Rev. Lett. **41**, 554 (1978).
[8] A. R. Bell, Phys. Plasmas **1**, 1643 (1994).
[9] R. J. Kingham and A. R. Bell, Phys. Rev. Lett. **88**, 045004 (2002).